\documentclass[12pt,a4paper,final]{iopart}
\usepackage{iopams}
\usepackage{float}
\usepackage{graphicx}
\usepackage[breaklinks=true,colorlinks=true,linkcolor=blue,urlcolor=blue,citecolor=blue]{hyperref}
\usepackage{tabularx}
\makeatletter

\newcommand{\Rmnum}[1]{\expandafter\@slowromancap\romannumeral #1@}

\makeatother

\begin{document}

\title{Elastic properties of a Sc-Zr-Nb-Ta-Rh-Pd high-entropy alloy superconductor}

\author{Yupeng Pan$^{a}$, Xiaobo He$^{a}$, Binjie Zhou$^{a}$, Denver Strong$^{b}$, Jian Zhang$^{a}$, Hai-Bin Yu$^{a}$, Yunfei Tan$^{a}$, Robert J. Cava$^{b*}$, and Yongkang Luo$^{a\dag}$}

\address{$^{a}$ Wuhan National High Magnetic Field Center and School of Physics, Huazhong University of Science and Technology, Wuhan 430074, China}
\address{$^{b}$ Department of Chemistry, Princeton University, Princeton, New Jersey 08544, United States}

\ead{rcava@princeton.edu; mpzslyk@gmail.com}

\date{\today}

\begin{abstract}
We report a comprehensive study on the elastic properties of a hexanary high-entropy alloy superconductor (ScZrNbTa)$_{0.685}$[RhPd]$_{0.315}$ at room and cryogenic temperatures, by Resonant Ultrasound Spectroscopy experiments. The derived elastic constants are bulk modulus $K=132.7$ GPa, Young's modulus $E=121.0$ GPa, shear modulus $G=44.9$ GPa, and Poisson's ratio $\nu$=0.348 for room temperature. The Young's and shear moduli are $\sim 10\%$ larger than those in NbTi superconductor with similar $T_c$, while the ductility is comparable. Moreover, the mechanical performance is further enhanced at cryogenic temperature. Our work confirms the advantageous mechanical properties of high-entropy alloy superconductors and suggests the application prospects.\\

\hspace{-15pt}\textbf{Keywords:} High-entropy alloy, Superconductor, Elastic constants

\end{abstract}

%74.25.Ld Mechanical and acoustical properties, elasticity, and ultrasonic attenuation
%62.20.Dc Elasticity, elastic constants
%43.35.Cg Ultrasonic velocity, dispersion, scattering, diffraction, and attenuation in solids; elastic constants

\maketitle

\section{Introduction}

Superconductors (SCs) are a class of ever-green functional materials discovered in 1911\cite{Onnes-SC1911}, with great application prospects in electrical current transfer, ground transportation, nuclear magnetic resonance (NMR) and medical resonance imaging (MRI), International Thermonuclear Experimental Reactor (ITER) and new-generation quantum computation as well. Until now, the so-called ``low-$T_c$ superconductors" Nb-Ti and Nb$_3$Sn based superconducting wires still dominate most of the commercial applications for superconducting eletromagnets. For industrial applications, the most crucial prerequisites of SC are high critical temperature ($T_c$), high upper critical field ($H_{c2}$), and high critical current density ($j_c$), which guarantee the strong magnetic field output. Apart from these factors, another technical issue one has to confront with is the mechanical performance. On the one hand, materials with good ductility are prone to make into wires; on the other hand, the strain effect exerted by the electromagnetic force on the superconducting wires reduces their superconducting properties\cite{Ekin-Strain}, and this is particularly the case for the A15-phase Nb$_3$Sn\cite{Ekin-Nb3SnStrain,Haken-Nb3SnStrain}. For these reasons, SCs with both high strength and good ductility are of considerable interest.

High-entropy alloys (HEAs) refer to systems containing more than four metallic elements in equimolar or near-equimolar ratios, offering a rich platform for materials design. Earlier studies on HEAs have revealed a series of intriguing properties, such as high hardness and strength\cite{Amar-Strength,Chen-Hardness}, simultaneous strength and ductility\cite{Zou-2015}, outstanding corrosion and oxidation resistance\cite{Chen-Hardness,Gorr-Oxidation,Zhang-Corrosion}, elegant strength-to-weight ratio\cite{Khaled-Light}, improved mechanical properties at cryogenic temperatures\cite{Gludovatz-Cryogenic}, \textit{etc}. A natural question concerns whether applicable superconducting wires can
be made of HEA. Indeed, HEA SCs present a unique crossing point between novel superconductors and functional high-entropy alloys that have attracted extensive interests in recent years.
In 2014, Ko\v{z}elj \textit{et al} reported the synthesis of the first HEA SC Ta$_{34}$Nb$_{33}$Hf$_8$Zr$_{14}$Ti$_{11}$ with $T_c\approx 7.3$ K\cite{Kozelj-TaNbHfZrTi}. One salient feature of this SC is that it exhibits extraordinarily robust zero-resistance superconductivity under pressure up to 190.6 GPa\cite{Guo-TaNbHfZrTiPres}. Later on, superconductivity was also observed in other HEAs \textit{e.g.} CsCl-type Sc-Zr-Nb-Rh-Pd, Sc-Zr-Nb-Ta-Rh-Pd\cite{Stolze-HEASC2018}, $Tr$Zr$_2$-type (Fe,Co,Ni,Rh,Ir)Zr$_2$ \cite{Mizuguchi-CuAl2HEASC,Kasem-TrZr2HEASC}, and \textit{hcp}-structured (MoReRu)$_{(1-2x)/3}$3(PdPt)$_x$ \cite{Zhu-MoReRuPdPt}. Most interestingly, the pentanary HEA SC (ScZrNb)$_{0.65}$[RhPd]$_{0.35}$ has $T_c\approx 9.7$ K and $H_{c2}\approx10.7$ T, comparable to those in NbTi\cite{Berlincourt-NbTi,Charifoulline-NbTi}, the superconducting alloy that accounts for a majority of the global superconductivity market for prevalent MRIs. Although HEA SCs have manifested themselves with excellent superconductivity, little has been known about their mechanical performance\cite{Zhu-MoReRuPdPt,Cheng-TiZrHfNbTafilm}, and in particular, a comprehensive study about their elastic constants and moduli at cryogenic temperature is still lacking.

Herein, by employing the Resonant Ultrasound Spectroscopy (RUS) technique, we performed a comprehensive study on the the second-rank elastic tensor of a representative HEA SC (ScZrNbTa)$_{0.685}$[RhPd]$_{0.315}$, which becomes a SC below $\sim$7 K. The derived elastic constants at room temperature are bulk modulus $K=132.7$ GPa, Young's modulus $E=121.0$ GPa, shear modulus $G=44.9$ GPa, and Poisson's ratio $\nu=0.348$. In particular, $E$ and $G$ are $\sim 10\%$ larger than the NbTi SC, while their $\nu$s are at the same level. These parameters suggest that this superconducting HEA possess both good strength and ductility. Meanwhile, unlike Nb$_3$Sn, the excellent mechanical performance retains even at cryogenic temperature. Our work confirms the advantageous mechanical properties of high-entropy alloy superconductors and suggests the application prospects.

\section{Experimental}

The polycrystalline hexanary HEA (ScZrNbTa)$_{0.685}$[RhPd]$_{0.315}$ sample studied here was grown by the arc-melting method as described elsewhere\cite{Stolze-HEASC2018}. The composition and structural characterizations were performd by X-ray diffraction (Cu-$K_{\alpha}$) and energy-dispersive X-ray spectroscopy (EDS) measurements. Electrical resistivity was measured as a function of temperature by the standard four-lead method in a commercial Physical Property Measurements System (PPMS-9, Quantum Design). For RUS measurements, the sample was carefully polished into a parallelepiped with the dimensions $1.152\times0.975\times0.586$ mm$^3$ and mass 6.33 mg. A schematic of the RUS experimental set-up is shown in Fig.~1(a). The transducers are made of a Lead Zirconate Titanate (PZT) plate and an Al$_2$O$_3$ hemisphere, and the latter is used for electrical isolation and mechanical protection\cite{LuoY-MnSiCreep}. A pair of transducers were used in RUS measurements, the bottom one as the ultrasound driving source, and the top one as the signal pick-up. To reduce the damping of the vibration modes, the sample is point-touch mounted between the two transducers. The measurements were made by sweeping frequency at fixed temperatures. More details about the RUS measurements can be found in Ref.~\cite{Migliori-RUS}. To measure the low-temperature elastic constants, a helium-flow cryostat (OptistatCF, Oxford) was exploited to cool the sample down to $\sim$5.4 K.

\begin{figure}[htbp]
%\hspace*{-14pt}
\includegraphics[width=9.5cm]{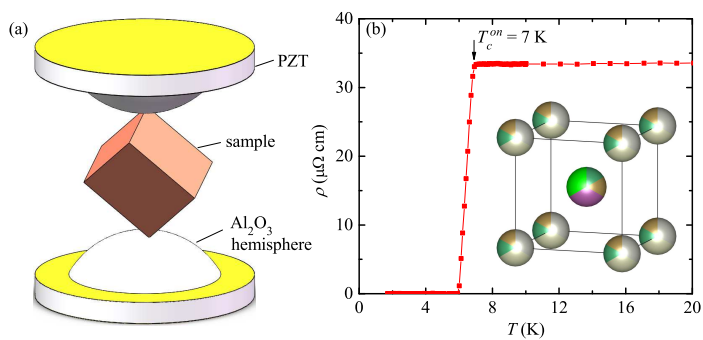}
%\vspace*{-15pt}
\label{Fig1}
\caption{(a) Schematic of RUS experimental set-up. (b) Temperature dependent resistivity of (ScZrNbTa)$_{0.685}$[RhPd]$_{0.315}$ showing an onset of the superconducting transition at $T_{c}^{on}=7$ K. The inset is the crystalline structure.}
\end{figure}

\section{Results and Discussion}

The composition of the sample studied in this paper is (ScZrNbTa)$_{0.685}$[RhPd]$_{0.315}$, which is a hexanary HEA superconductor with onset superconducting transition $T_c^{on}\approx7$ K, verified by resistivity measurements shown in Fig.~1(b). Bulk nature of superconductivity was confirmed by the Meissner effect in our previous work\cite{Stolze-HEASC2018}. This material has CsCl-type structure and mixed-site occupancies\cite{Stolze-HEASC2018}. The advantage of RUS experiment is that it can extract the full elastic tensor $\{C_{ij}\}$ ($i,j$=1-6) in a single frequency sweep. Because the sample is an isotropic polycrystal, the elastic tensor has only two independent elements, viz. $C_{11}$ and $C_{44}$ in Voigt notation, whereas $C_{12}$ can be retrieved by $C_{12}=C_{11}-2C_{44}$.

\begin{figure}[htbp]
\hspace*{-0pt}
\includegraphics[width=9.5cm]{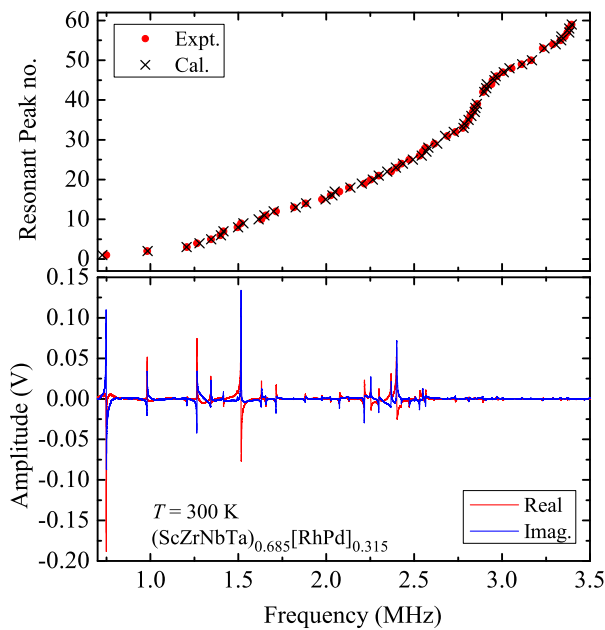}
%\vspace*{-30pt}
\caption{\label{Fig2}Bottom panel, vibrational spectrum of (ScZrNbTa)$_{0.685}$[RhPd]$_{0.315}$ at room temperature. Top panel shows the first 59 resonances, red circles - experimental data, and  black crosses - calculated data.}
\end{figure}

For a sample with given elastic constants, density and dimensions, theoretically, all the normal resonance modes can be computed directly by solving a three dimensional elastic wave function\cite{Migliori-RUS,Migliori-PhysicaB1993,Leisure-RUS}. The way we did for RUS measurements is opposite. We swept frequency from 0.7 to 3.5 MHz, and collected the full spectrum at different temperatures. A representative spectrum is displayed in the bottom panel of Fig.~2. About 59 resonant peaks can be recognized in this range, and each stands for a specific vibration mode. The elastic constants are derived by a computerized fitting algorithm with a least-square criterion, in which $C_{11}$ and $C_{44}$ are set as free fitting parameters. The iteration continues until $\chi^2\equiv\sum_n[(F_n^{expt}-F_n^{cal})/F_n^{expt}]^2$ minimizes, where $F_n^{cal}$ and $F_n^{expt}$ are the $n$th calculated and experimental frequencies, respectively. The fitting yields $C_{11}=192.6$ GPa, $C_{12}=102.8$ GPa, and $C_{44}=44.9$ GPa for $T=300$ K. The fitting pattern is presented in the top panel of Fig.~2. The shear modulus $G=C_{44}=44.9$ GPa is obtained directly.

With these elastic parameters, we further calculate the bulk modulus \cite{Migliori-RUS}
\begin{equation}
    K=\frac{C_{11}+2C_{12}}{3}=132.7~GPa,
    \label{Eq1}
\end{equation}
Young's modulus
\begin{equation}
    E=\frac{9KG}{3K+G}=121.0~GPa,
    \label{Eq2}
\end{equation}
and Poisson's ratio
\begin{equation}
    \nu=\frac{3K-2G}{2(3K+G)}=0.348.
    \label{Eq3}
\end{equation}

According to Pugh\cite{Pugh-Plastic}, materials with the ratio $K/G>$1.75 are classified as plastic. For (ScZrNbTa)$_{0.685}$[RhPd]$_{0.315}$, this ratio reaches 2.95, indicative of good plasticity. This is further supported by the Poisson's ratio. $\nu$ of covalent systems are known to be small ($\nu\sim0.1$), while those of ionic crystals are $\sim0.25$\cite{Shein-FeAsDFT}. The value $\nu=0.348$ in (ScZrNbTa)$_{0.685}$[RhPd]$_{0.315}$ resides in a plastic regime ($\nu>0.31$)\cite{Lewandowski-Plastic}.

\begin{table}[!ht]
\tabcolsep 0pt \caption{\label{Tab.1} Elastic constants of representative conductors and superconductors at room temperature. $K$ - Bulk modulus, $E$ - Young's modulus, $G$ - Shear modulus, $\nu$ - Poisson's ratio. Acoustic Debye temperature $\Theta_D$ is calculated according to the elastic constants.}
\vspace*{-12pt}
%\begin{center}
\def\temptablewidth{1.1\columnwidth}
{\rule{\temptablewidth}{1pt}}
\begin{tabular*}{\temptablewidth}{@{\extracolsep{\fill}}ccccccc}
Materials   &  $T_c$ (K)  & $K$ (GPa)   &    $E$ (GPa)  &   $G$ (GPa)     & $\nu$     &  $\Theta_D$  (K)  %  &  $\kappa_L$ (W/m$\cdot$K)
\\ \hline
(ScZrNbTa)$_{0.685}$[RhPd]$_{0.315}$ & 7    & 132.7   &  121.0  &  44.9  &  0.348   &  277  %&   13
\\
Cu\cite{HM-Cu2009}                   & $-$  & 140.2   &  123.5  &  45.4  &  0.350   &  332  %&   15
\\
Nb\cite{Carroll-Nb}                  & 9.2  & 174.3   &  108.9  &  39.0  &  0.396   &  274  %&   7
\\
NbTi\cite{HM-NbTi2008}               & 9.7  & 131.8   &  111.7  &  41.1  &  0.359   &  325  %&   14
\\
Nb$_3$Sn\cite{Rehwald-Nb3Sn}         & 18   & 155.6   &  139.4  &  51.6  &  0.351   &  305  %&   26
\\
V$_3$Ga\cite{Bussiere-V3Ga}          & 15   &         &  115.0  &        &          &       %&
\\
YNi$_2$B$_2$C\cite{Rourke-YNi2B2CRUS}& 15.3 & 184.7   &  218.1  &  83.7  &  0.303   &  553  %&   88
\\
MgB$_2$\cite{Nesterenko-MgB2}        & 39   & 128.0   &  245.0  &  104.0 &  0.180   &  971  %&  325
\\
YBCO\cite{Lei-OxideSC}               & 90   & 115.9   &  167.9  &  66.7  &  0.259   &  462  %&  126
\\
LaFeAsO$^\dag$\cite{McGuire-LaFeAsO} & $-$   & 47.3    &   73.9  &  29.8  &  0.240   &  300  %&   46
\\
\end{tabular*}
{\rule{\temptablewidth}{1pt}}
%\end{center}
\begin{flushleft}
$^\dag$ LaFeAsO becomes a superconductor when electron-doped by partially substituting O with F, and the $T_c$ of LaFeAsO$_{0.89}$F$_{0.11}$ reaches 26 K\cite{Kamihara-La1111_F}. The elastic moduli of LaFeAsO derived from powder polycrystal are probably underestimated. The values from first-principles calculation are\cite{Shein-FeAsDFT}: $K=97.9$ GPa, $E=141.5$ GPa, $G=56.2$ GPa, $\nu=0.259$, $\Theta_D=416$ K.%, $\kappa_L=81$ W/m$\cdot$K.     \\
\end{flushleft}
\end{table}

The optimized elastic constants from RUS measurements also enable us to estimate some thermodynamic parameters. The Debye temperature ($\Theta_D$) is related to the average sound velocity ($v_m$) via \cite{Anderson-Debye}
\begin{equation}
    \Theta_D=\frac{h}{k_B}[\frac{3n\rho N_A}{4\pi M}]^{1/3}v_m,
    \label{Eq4}
\end{equation}
where $h$ and $k_B$ are as conventionally defined in quantum mechanics, $N_A$ is Avogadro's number, $\rho$=9.610 g/cm$^3$ is the density, $M$ is the molecular weight of the solid, and $n$=2 is the number of atoms in the CsCl-type molecule\cite{Anderson-Debye}. Here we adopted the average molecular weight $M$=206.376 g/mol for (ScZrNbTa)$_{0.685}$[RhPd]$_{0.315}$. The average sound velocity is taken as \cite{Jamal-Cubic}
\begin{equation}
    \frac{1}{v_m^3}=\frac{1}{3}(\frac{1}{v_{11}^3}+\frac{2}{v_{44}^3}),
    \label{Eq5}
\end{equation}
where $v_{11}=4477$ m/s and $v_{44}$=2162 m/s are the longitudinal and shear sound velocities, respectively, and they can be retrieved from $v_{ii}=\sqrt{C_{ii}/\rho}$ ($i$=1,4). The calculations result in $v_m=2430$ m/s and $\Theta_D=277$ K.

In addition, the lattice thermal conductivity $\kappa_L$ (for $T>\Theta_D$) can be evaluated by\cite{Slack-Thermocond,Slack-1979}
\begin{equation}
    \kappa_L=\frac{A\bar{M}\delta n^{1/3}\Theta_D^3}{\gamma^2 T},
    \label{Eq6}
\end{equation}
where $\bar{M}=103.188$ g/mol is the average mass of the atoms in the crystal, $\gamma\equiv\frac{3}{2}(\frac{3v_{11}^2-4v_{44}^2}{v_{11}^2+2v_{44}^2})=2.1$ is the acoustic Gr\"{u}neisen parameter that characterizes the anharmonicity of a sample \cite{Belomestnykh-PoissionGruneisen,Jia-LattKappa}, $A=\frac{2.43\times10^{4}}{1-0.514/\gamma+0.228/\gamma^2}$ W/m$^2\cdot$K$^3$, and $\delta^3$ signifies the average volume occupied by one atom in the crystal that can be known from the structural parameters\cite{Stolze-HEASC2018}. This gives rise to $\kappa_L=16.0$ W/m$\cdot$K at 300 K.

The elastic constants and the estimated thermodynamic parameters for (ScZrNbTa)$_{0.685}$[RhPd]$_{0.315}$ are summarized in Table \ref{Tab.1}, and for comparison, we also list other representative conductors and superconductors. It is interesting to note that the Young's and shear moduli of (ScZrNbTa)$_{0.685}$[RhPd]$_{0.315}$ are about 10\% larger than in Nb and NbTi that have comparable $T_c$. Meanwhile, the Poisson's ratios of (ScZrNbTa)$_{0.685}$[RhPd]$_{0.315}$, Cu and NbTi are essentially the same, implying that they have similar ductility. Therefore, filamentary superconducting wires made of (ScZrNbTa)$_{0.685}$[RhPd]$_{0.315}$ / Cu composite are possible.

\begin{figure}[htbp]
\hspace*{-0pt}
\includegraphics[width=9.5cm]{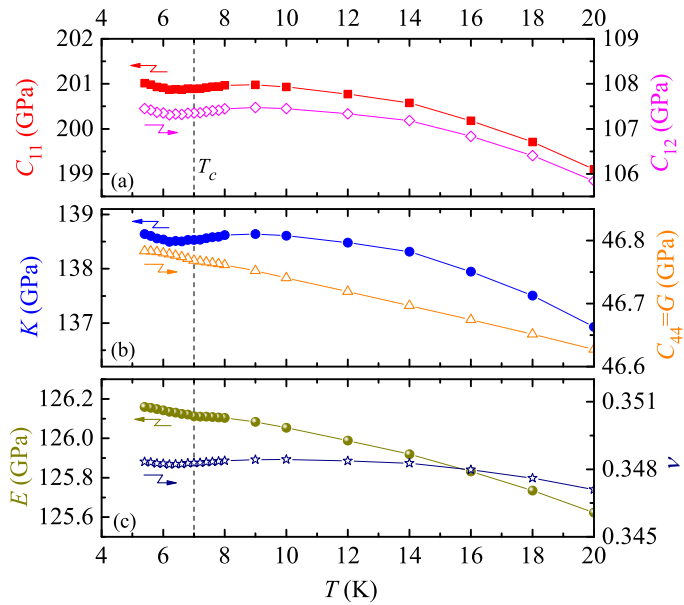}
%\vspace*{-30pt}
\label{Fig3}
\caption{Low-temperature elastic constants of (ScZrNbTa)$_{0.685}$[RhPd]$_{0.315}$. (a), $C_{11}$ (left) and $C_{12}$ (right); (b), $K$ (left) and $G$ (right); (c) $E$ (left) and $\nu$ (right). The dashed lines mark $T_c$.}
\end{figure}

In order to study the mechanical performance under cryogenic conditions, we also performed the RUS measurements at low temperature, and the results are shown in Fig.~3. Upon cooling, $C_{44}$ ($=G$) increases monotonically below 20 K, and a weak inflection is visible at $T_c$, manifesting the superconducting transition. The same trend is also seen in Young's modulus. Other elastic moduli including $C_{11}$, $C_{12}$ and $K$ also increase below 20 K, and tend to level off, but soften slightly near $T_c$. The tiny feature of elastic constants at $T_c$ probably implies relatively weak electron-phonon coupling, which is conceivable here. In particular, the absence of a step-like discontinuity in $C_{ij}$ around $T_c$ evidences that the coupling between strain and superconducting order parameter is very weak\cite{Luthi-Acoustics}. At 5.4 K, the base temperature of our measurements, $K=138.6$ GPa, $E=126.2$ GPa, and $G=46.8$ GPa. It is important to note that for all the temperatures tested, Poisson's ratio remains essentially constant about 0.348. All these suggest that at cryogenic condition this HEA exhibits even better mechanical performance than at room temperature. We should also point out that in Nb$_3$Sn, due to the formation of martensitic phase ($\sim 43$ K), Young's and shear moduli are softened dramatically, and the values reduce to $E=49$ GPa and $G=16.8$ GPa at 4.2 K\cite{Poirier-Nb3Sn}. This makes Nb$_3$Sn rather brittle and in turn causes the large strain dependence in the critical current density\cite{Ekin-Nb3SnStrain,Haken-Nb3SnStrain}. Such a shortcoming is absent in (ScZrNbTa)$_{0.685}$[RhPd]$_{0.315}$.

Finally, it is important to note that among the SCs listed in Table 1, (ScZrNbTa)$_{0.685}$[RhPd]$_{0.315}$ has the elastic constants most close to Cu; in other words, the elastic properties of (ScZrNbTa)$_{0.685}$[RhPd]$_{0.315}$ and Cu are in good compatibility. This suggests that the filamentary superconducting wire made of (ScZrNbTa)$_{0.685}$[RhPd]$_{0.315}$ / Cu subjects little strain effect, and thus will maintain good superconductivity. Also, because all the elastic moduli of Cu are relatively larger than in (ScZrNbTa)$_{0.685}$[RhPd]$_{0.315}$, the (ScZrNbTa)$_{0.685}$[RhPd]$_{0.315}$ / Cu composite wire is expected to exhibit even better mechanical performance than pure (ScZrNbTa)$_{0.685}$[RhPd]$_{0.315}$.

\section{Conclusion}

In conclusion, the elastic properties of the high-entropy alloy superconductor (ScZrNbTa)$_{0.685}$[RhPd]$_{0.315}$ have been studied by Resonant Ultrasound Spectroscopy measurements. The room-temperature bulk modulus is $K=132.7$ GPa, Young' modulus is $E=121.0$ GPa, shear modulus is $G=44.9$ GPa, and Poisson's ratio is $\nu=0.348$. The Young's and shear moduli are $\sim$10\% larger than those in NbTi superconductor with similar $T_c$. Most crucially, the mechanical performance is further improved at cryogenic temperature. These results illustrate the advantageous elastic properties of high-entropy alloy superconductors, and suggest the feasibility for industrial applications.

\section{CRediT authorship contribution statement}

\textbf{Yupeng Pan:} Elastic property measurements, Data analysis, Writing. \textbf{Xiaobo He:} Resistivity measurements. \textbf{Binjie Zhou:} Elastic property measurements. \textbf{Denver Strong:} Sample synthesis. \textbf{Jian
Zhang:} Elastic property measurements. \textbf{Hai-Bin Yu:} Validation, Guidance. \textbf{Yunfei Tan:} Validation, Guidance. \textbf{Robert J. Cava:} Sample synthesis, Validation, Guidance. \textbf{Yongkang Luo:} Supervision, Methodology, Writing.

\section{Declaration of Competing Interest}

The authors declare that they have no known competing financial interests or personal relationships that could have appeared to influence the work reported in this paper.

\section{Acknowledgments}

We thank Albert Migliori and Brad Ramshaw for technical aid. This work was supported by National Natural Science Foundation of China (No. 52077086), and the sample preparation at Princeton University was supported by the US Department of Energy Division of Basic Energy Sciences grant number DE-FG02-98ER45706.

\section{Author contributions}

The manuscript was written with contributions from all the authors. All the authors have approved the final version of the manuscript.

\section*{References}

%\bibliographystyle{iopart-num}
%\bibliography{biblio}
%\bibliographystyle{ieeetr}
%\bibliographystyle{iopart-num}

\providecommand{\newblock}{}

\end{document}